\documentclass[runningheads]{svmult}
\usepackage{makeidx}   
\usepackage{graphicx}  
\usepackage{subeqnar}  
\usepackage{multicol}  
\usepackage{cropmark} 
\usepackage{physprbb}  
\usepackage{amsmath}
\usepackage{amssymb}

\newcommand{\hetrois}    {\mbox{$ ^{3}{\mathrm{He}}                            $}~}
\newcommand{\hetro}    {\mbox{$ ^{3}{\mathrm{He}}                            $}}
\newcommand{\tritium}    {\mbox{$ ^{3}{\mathrm{H}}                            $}~}
\newcommand{\neut}{$\tilde{\chi}$~}
\newcommand{\neutt}{$\tilde{\chi}$}

\def\PRB#1#2#3{{\rm Phys. Rev.} {\bf{B#1}} (19#2) #3}
\def\PRL#1#2#3{{\rm Phys.~Rev.~Lett.} {\bf{#1}} (19#2) #3}
\begin{document}
\title*{A project of a new detector for direct Dark Matter search: MACHe3}
\toctitle{A project of a new detector for direct Dark Matter search: MACHe3}
%
%
\titlerunning{MACHe3 project}
%
\author{D.Santos\inst{1}
\and F.Mayet\inst{1}
\and Yu.M.Bunkov\inst{2}
\and A.Drezet\inst{1}
\and G.Duhamel\inst{1}
\and H.Godfrin\inst{2}
\and F.Naraghi\inst{1}
\and G.Perrin\inst{1}}
\authorrunning{D.Santos et al.}
%
%
\institute{Institut des Sciences Nucl\'eaires,
CNRS/IN2P3 and Universit\'e Joseph Fourier,\\
 53, avenue des Martyrs, 38026 Grenoble cedex, France
\and Centre de Recherches sur les Tr\`es Basses Temp\'eratures, \\
 CNRS, BP166, 38042 Grenoble cedex 9, France\\}

\maketitle              

\begin{abstract}
MACHe3 (MAtrix of Cells of superfluid \hetro) is a project of a new detector 
for direct Dark Matter (DM) search. A cell of superfluid \hetrois
has been developed and the idea of using a large number of such cells in a 
high granularity detector is proposed.This contribution presents, after a brief description of the superfluid 
\hetrois cell, the simulation of the response 
 of different matrix configurations allowing to define an optimum design 
 as a function of the number of cells and the volume
 of each cell. The exclusion plot and the predicted interaction cross-section for the
 neutralino as a photino  are 
 presented.
\end{abstract}

\section{Introduction}
In the last two decades many experiments on direct detection of DM have been
performed. Many of them are  running today providing new results which
 have imposed upper limits for the interaction cross-section of 
these particles with the stable baryons : neutrons and protons. They use a great
variety of materials and detection techniques. However the neutrons are still very difficult to 
discriminate along with a good rejection for gamma rays coming from the natural
radioactivity. The cosmogonic activation is in all these experiments an important
source of intrinsic contamination.

In general the sensitive medium of an ideal detector of WIMPs should 
have the following desired properties :
\begin{itemize}
\item  to be composed of odd nuclei in order to have as well
as the scalar interaction (coherent) the axial spin-spin interaction with the 
hypothetical particles

\item  to present a high neutron capture cross-section to sign the neutron background

\item  to be produced with a high purity, mainly free of radioactive isotopes

\item  to present a Compton interaction cross-section as low as possible in order 
to minimize the
interaction with the natural gamma radioactivity background

\end{itemize}

\noindent
In addition, the detection device should allow to get a threshold energy as low as 
possible.
\hetrois in the superfluid phase B at ultra low temperatures (T $\simeq$ 100 
$\mu$ K) presents in the Lancaster's configuration \cite{lanc}   a low detection
threshold ($\simeq $1 keV) in  a  ultra  high purity state. The other features are largely
fulfilled by the \hetrois nucleus itself. 

Nevertheless to further enhance  the background rejection we propose
a matrix configuration in which the correlations among cells allow to improve
the discrimination of background events. This rejection has been confirmed and
quantitatively estimated by a Monte Carlo simulation study described below. 

\section{Description of a cell}
The primary device consisted of a small copper cubic box (V$\simeq$ 0.125 cm$^{3}$)
 filled with \hetro. It is immersed in a larger
volume containing liquid \hetrois and thin plates of copper nuclear-cooling 
refrigerant, see fig.1. 
Two vibrating wires are placed inside the cell, forming a Lancaster type bolometer
\cite{prl95}. A small hole on one 
of the box walls connects the box to the main \hetrois volume, 
thus allowing the diffusion of the thermal excitations of the \hetrois generated 
by the energy deposited in the bolometer by the interacting particle.\\
This high sensitivity device is used as follows : the incoming particle deposits an 
amount of energy in the cell, which is converted into 
 quasiparticles of the superfluid state. These are detected by their damping effect
  on the vibrating wire. It must be pointed out that the size of the
hole governs the relaxation time (quasiparticles escape time) and the Q factor of the 
resonator governs the rising time.  
 The present device has a rather high Q factor (Q $\simeq 10^{4}$), giving a rising 
 time of the order of one second.
Although the primary experiment \cite{prl95} was still rudimentary, it has allowed to detect 
signals down to a threshold of 1 keV.
Many ideas are under study to improve the sensitivity of such a cell. Recently, the fabrication of 
micromechanical silicon resonators has been reported \cite{trique} and the possibility to use such wires at
ultra-low temperatures is under study.\\

\begin{figure}[htb]
\begin{center}
\includegraphics[width=.6\textwidth]{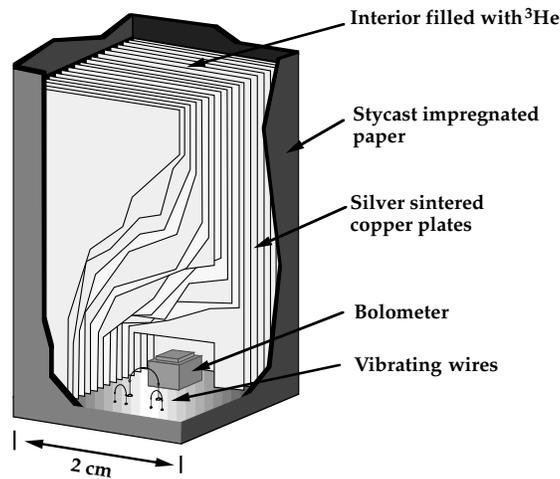}
\end{center}
\caption[]{Ultra-low temperature nuclear stage and bolometer cell}
\label{eps1.1}
\end{figure}
\newpage
\section{Simulation of a high granularity \hetrois detector.}
The performance of a detector for direct DM search is based on the discrimination
capability of the different types of background events.
The main background components are : thermal and fast neutrons, muons and gamma rays.
The aim of the simulation described below was to evaluate the total rejection of
different configurations of matrices of \hetrois cells taking into account the
correlation among the cells and the energy loss in each cell. 

As a typical WIMP is expected to transfer up to 6 keV, it is necessary to evaluate the 
proportion of background events releasing less than 6 keV in the \hetrois cell.\\
The elastic cross-section between a WIMP and \hetrois being  small ($\sigma \lesssim 10^{-3} pb$), a
 WIMP event will be characterized by a single-cell event, with equal probability among all the cells of the matrix. 
Hence, the rejection against background events will be achieved by choosing events having the following characteristics :
\begin{itemize}
\item 
Only one cell fired (single-cell event). The quality parameter related to this 
selection will be referred to as  
Correlation Coefficient.
\item 
Energy released in this cell below 6 keV (Energy Rejection).
\item 
The outermost cell layer is considered as a veto (off-line), hence the fired cell has to be in the inner part of the matrix. 
This will reject neutrons interacting elastically.
\end{itemize}
\begin{figure}[htb]
\begin{center}
\includegraphics[width=.5\textwidth]{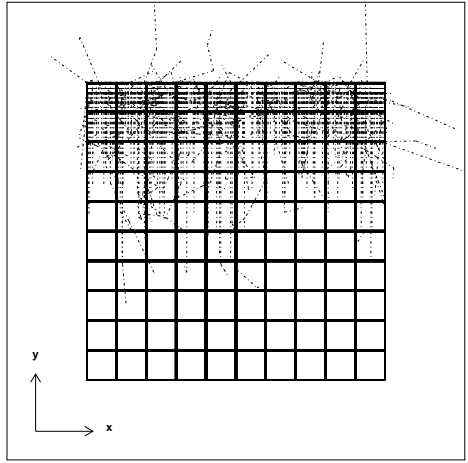}
\end{center}
\caption[]{2-dimensional view of a proposed matrix of 1000 cells (125 $cm^{3}$ each).
 The events,  generated in a direction 
perpendicular to the upper face, are 10 keV neutrons. It can be noticed that most of
neutrons of this energy are captured in the first layer.}
\label{10kev}
\end{figure}
A simulation has been done, in order to define an optimum design as 
a function of the number of cells and the volume of each cell. Then, the rejection capability has been
estimated, for the preferred design, as well as  the false WIMP rate.\\
The simulation has been done with a complete Monte-Carlo simulation using GEANT3.21 \cite{geant} package and 
in particular the GCALOR-MICAP(1.04/10) \cite{micap} package for slow neutrons.
The simulated detector (fig. \ref{10kev}) consists of a cube containing a variable number of cubic \hetrois cells, immersed 
in a large volume containing \hetrois ($\rho_{SF}$=0.08 $g.cm^{-3}$). Each cell is surrounded by a thin copper layer and it is 
separated from the others by a 2 $mm$ gap (filled with \hetro).\\

In contrast to most DM detectors, MACHe3 may be sensitive to rather low energy neutrons, and its response depends strongly on their kinetic
energies. The total cross-section interaction for a neutron in \hetrois ranges from $\sigma_{tot} \simeq 1000$ barns, 
for low energy neutrons(E$_{n} \simeq$ 1 eV), down to $\sigma_{tot} \simeq 1$ barn
for 1 MeV neutrons. The main processes are : elastic scattering which starts being predominant above 600 keV, and neutron 
capture: \hetrois (n,p) \tritium , which is largely predominant for low 
energy neutrons ($E_{n} \leq 10$ keV) :

\begin{equation}
n+ ^{3}He \rightarrow p+^{3}H +764 keV
\end{equation}

\begin{figure}[hbt]
\begin{center}
\includegraphics[width=.49\textwidth]{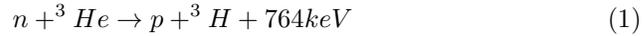}
\includegraphics[width=.49\textwidth]{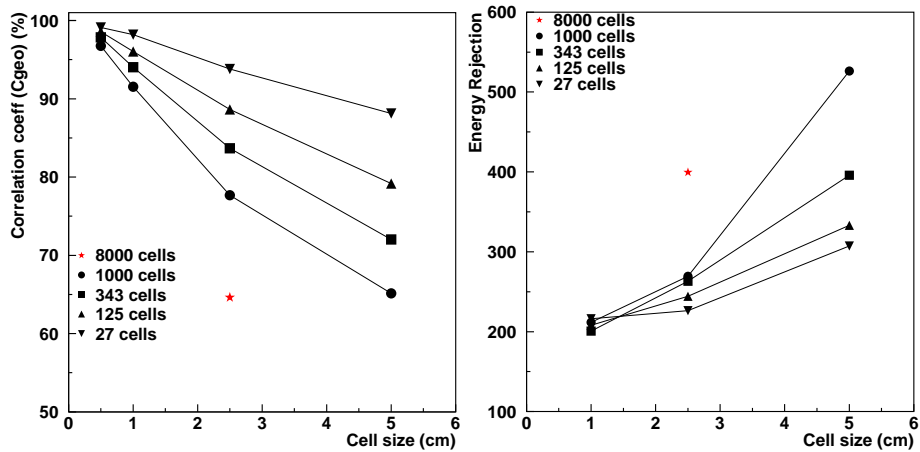}
\end{center}
\caption[]{Correlation Coefficient and Energy Rejection as a function of the size of the \hetrois cell for 1 MeV neutrons. 
The different curves correspond to different matrix sizes as indicated by the labels.}
\label{coeff}
\end{figure}
\noindent
The energy released by the neutron capture is shared by the recoil ions : the tritium $^{3}$H with kinetic energy 191 keV 
and the proton with kinetic energy 573 keV. The range for these two particles
is fairly short : typically 12 $\mu$m for tritium and 67 $\mu$m for proton; consequently neutrons undergoing capture in \hetrois are expected 
to produce 764 keV within the cell, 
thus being clearly separated from the expected \neut signal (E $\leq $ 6keV). The tritium produced by neutron capture
will eventually decay with a half-life of 12 years by $\beta$-decay with an end-point electron spectrum at 18 keV. It 
means that the number of neutrons capture per cell must be counted to estimate the contribution of this kind of events
on the false \neut rate.\\
The neutron capture cross-section decreases with increasing neutron kinetic energy, but on the other hand, the energy released in 
the \hetrois cell by the elastic scattering is getting larger, thus
diminishing the probability to leave less than 6 keV.

\begin{figure}[htb]
\begin{center}
\includegraphics[width=.6\textwidth]{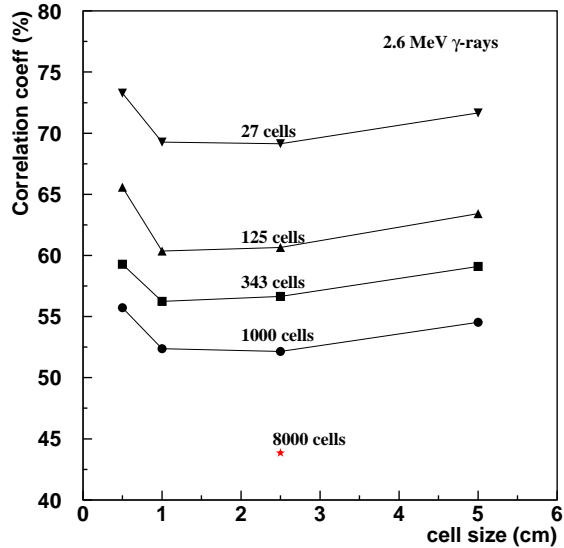}
\end{center}
\caption[]{Correlation coefficient as a function of the size of the \hetrois cell
for 2.6 MeV $\gamma $-rays. The different curves correspond to different matrix 
sizes as indicated by the labels.}
\label{gam2.6}
\end{figure}

We present here, the main quality parameters for 1 MeV neutrons (fig. \ref{coeff}) and
for
2.6 MeV $\gamma$-rays (fig. \ref{gam2.6}).
This coefficients \footnote{With the definition of the quality parameters \cite{firstmac3}, the best design is the one for which 
the correlation coefficient is the lowest (thus minimizing the proportion of single-cell background events) 
and the energy rejection is the highest (meaning a low proportion of background events with an energy measurement below 6 keV).} 
depend both on the matrix and cell sizes. The full simulation for  various kinetic
energies has  
shown \cite{firstmac3} that a large cell (125 $cm^3$) allows to obtain a large energy rejection (R$\sim \! 500$ for 1 MeV neutrons), 
and a large matrix (1000 cells or more) allows to have a good correlation among the cells ($\sim$ 65\% for 1 MeV neutrons), 
thus rejecting efficiently $\gamma$-rays and neutrons. For background rejection consideration the optimum configuration 
is a matrix of 1000 cells of 5 $cm$ side (more details may be found in \cite{firstmac3}
).

\begin{figure}[htb]
\begin{center}
\includegraphics[width=.6\textwidth]{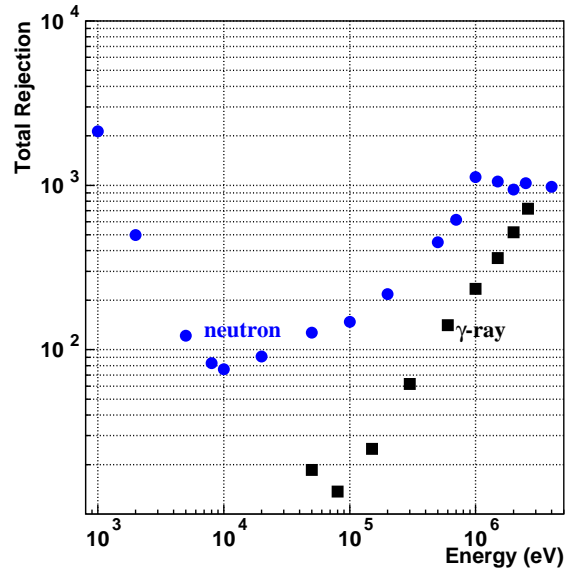}
\end{center}
\caption[]{Total Rejection as a function of the incident particle energy, for a matrix of 1000 cells (125
$cm^{3}$ each). The different set of points correspond to $\gamma$-rays (squares) and neutrons (circles). It must be pointed out that the evaluated rejection is for a "naked matrix", i.e. without taking into account any lead or paraffin shielding
 or any separation between electron and ion recoils. It represents the capability of the \hetrois matrix to reject background events by means of energy loss
measurements and correlation considerations.}
\label{rejec}
\end{figure}

A detailed simulation has been done for the matrix of 1000 cells of 5 $cm$ side 
representing a good 
compromise between rejection power, volume  and feasibility. 
For neutrons, a large rejection\footnote{ The total rejection is defined as the ratio 
between the number of incoming particles and the number of false \neut events 
(less than 6 keV in one non-peripheric cell).} is achieved below 8 keV (mainly thanks to veto and energy rejection) and above 1 MeV (thanks to
correlation and energy loss measurement). As expected, 8 keV neutrons represent the worst rejection (see fig. \ref{rejec}), 
since the capture process is less predominant, and the energy left by (n,n) 
interaction is always less than 6 keV.\\
\noindent
For gamma rays, a high granularity  detector provides an intrinsic rejection ranging 
between 10 and 1000, depending on their kinetic energies. This selection, may be 
improved by adding a discrimination between
recoils and electrons under study. Different experimental approaches should be tested 
shortly. A complete study of an inner and outer cryostat shielding is also
needed, as well as an evaluation of natural radioactivity of materials.\\

As neutrons recoiling off nuclei may easily simulate a \neut event, 
it is crucial to evaluate the neutron-induced false event rate. 
 For this purpose, a simulation of a paraffin neutron shielding has also been done, in order to evaluate the expected 
neutron spectrum through this shielding. We have used the measured neutron spectrum \cite{Chazal:1998qn} in {\em Laboratoire Souterrain 
de Modane}, between 2 and 6 MeV \footnote{The thermal neutron flux, evaluated 
in \cite{Chazal:1998qn} to be (1.6$\pm0.1)\times 10^{-6} cm^{-2}s^{-1}$, will be highly suppressed by the 30 $cm$ paraffin shielding}, 
with an integrated flux of $\Phi_{n}\simeq 4\times 10^{-6} cm^{-2}s^{-1}$.\\
Using this flux and the expected rejection factor (fig. \ref{rejec}), we evaluated the false
\neut rate induced by neutron background, to be $\sim$0.1 false event per day through 
the 1.5$m^{2}$ surface of the detector 
(1000 cells of 125 $cm^{3}$), which is  much lower than the expected \neut rate (of the order of $\sim$1 day$^{-1}$
in a detector of this size, see below and \cite{next}). 

\noindent
A study of the muon background has also been done. We estimated \cite{firstmac3} the $\mu$-induced false \neut rate to be of the order 
of 0.01 day$^{-1}$m$^{-2}$, which is more than two orders of magnitude 
below the expected \neut rate. 

\section {Cross-section and LSP }
\begin{figure}[htb]
\begin{center}
\includegraphics[width=.6\textwidth]{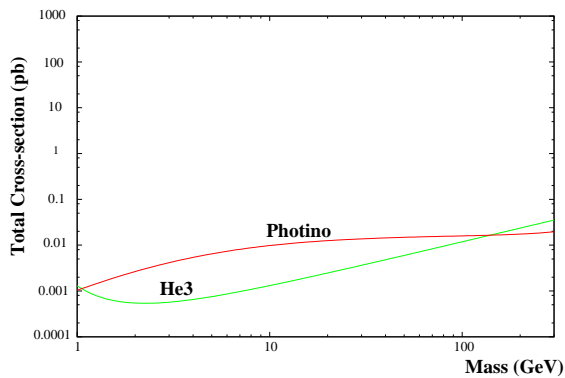}
\end{center}
\caption[]{Accessible total cross-sections $\tilde{\chi}$-\hetro, in pb, for
  10 kg of
\hetrois as a function of the neutralino mass. The curve showing the predicted
cross-section as a function of the neutralino mass for the special case of a pure
photino is also shown}
\label{lsp}
\end{figure}

 The plot showing the accessible cross-section interaction for a rate of 0.1
 events per day and per kg as a function of the mass of the WIMP assuming a Maxwellian distribution
 of the WIMP's velocities in the galactic halo is shown on fig. \ref{lsp} for  10 kg of 
\hetro. 

\noindent
However we have to verify if the available models allowing to estimate the
 interaction cross-section in the case of the best candidate for the lightest SUSY 
 particle (LSP) as the neutralino give a value accessible by this  plot. Doing that 
 for one of the special cases in which the neutralino is a pure photino we get an overlap
 shown on fig. \ref{lsp}. This special case is one example in which only the axial interaction
 channel is available to detect the elusive particle.
 

\section{Conclusion}
As a conclusion, it can been said that a large matrix ($\sim$ 1000 cells) of large cells (125 $cm^{3}$) is the
preferred design for a multi-cell superfluid \hetrois detector. By means of the 
correlation among the cells and the energy loss measurement, 
a high rejection may be obtained for $\gamma$-ray, neutron and muon background. 
For background rejection purpose, the main advantage of a superfluid \hetrois detector 
is to present a high rejection against neutron background, mainly because of the high 
capture cross-section at low energy;  
as neutrons interact {\it a priori} like \neutt, they are the ultimate background noise for DM detectors.\\

%
%
%

%

%
%

%

\end{document}